\documentclass[aip,twocolumn,reprint,preprintnumbers,amsmath,amssymb]{revtex4-1}
\usepackage{graphicx}
\usepackage{dcolumn}
\usepackage{bm}
\usepackage{amssymb}
\usepackage{epsfig}

\usepackage{epsfig}
\usepackage{amsmath}

\usepackage{amsfonts}
\usepackage{mathptmx}
\usepackage{eucal}

\usepackage{color}
\usepackage[colorlinks,linkcolor=blue,citecolor=blue]{hyperref}
\usepackage{epstopdf}

\usepackage{natbib}
\begin{document}


\title{Efficient phase-tunable Josephson thermal rectifier}

\author{M. J. Mart\'{\i}nez-P\'erez}
\email{mariajose.martinez@sns.it}
\affiliation{NEST, Istituto Nanoscienze-CNR and Scuola Normale Superiore, I-56127 Pisa, Italy}

\author{F. Giazotto}
\email{giazotto@sns.it}
\affiliation{NEST, Istituto Nanoscienze-CNR and Scuola Normale Superiore, I-56127 Pisa, Italy}




\begin{abstract}

Josephson tunnel junctions are proposed as efficient phase-tunable thermal rectifiers. The latter exploit the strong temperature dependence of the superconducting density of states and phase-dependence of heat currents flowing through Josephson junctions to operate. Remarkably, large heat rectification coefficients up to $\sim 800\%$ can potentially be achieved using conventional materials and standard fabrication methods. In addition, these devices allow for the \emph{in-situ} fine tuning of the thermal rectification magnitude and direction. 

\end{abstract}

\pacs{}

\maketitle



Electronic circuits consist of a number of components (e.g., transistors, diodes and switches) connected together to enable the execution of different operations. In metals, electrons are responsible of energy transport as well, in what is commonly referred to as electronic heat transport.\cite{GiazottoRev,Dubi} It is therefore natural to address the feasibility of networks that might eventually allow for the implementation of, for instance, thermal computation, thermal logic operations or data storage (for a review see Ref. \citenum{LiRev} and references therein). Additionally,  mastering of heat currents represents an important breakthrough in different research fields of nanoscience such as solid-state cooling,\cite{GiazottoRev} radiation detection\cite{GiazottoRev}, quantum computing\cite{NielsenChuang} or the emerging field of \emph{coherent caloritronics}.\cite{Giazottoarxiv,GiazottoAPL12,martinez} So far, a strong effort has been devoted to envision thermal rectifiers, i.e., structures allowing high heat conduction along one direction but suppressed thermal transport upon temperature bias reversal.\cite{RobertsRev,CasatiNandV} Most of these proposals deal with phononic heat transport,\cite{Li,Segal,Segal2,Terraneo} very few deal with electronic heat conduction\cite{ren,Ruokola1,Kuo,Ruokola2,Chen} and even less have demonstrated feasible experimental realizations.\cite{Scheibner,Chang}    

In this Letter we propose and analyze theoretically the performance of a thermal diode consisting of a SIS' Josephson tunnel junction, where the I stands for an insulating barrier and S and S' represent two different superconducting electrodes. Although never considered so far for such a purpose, superconducting tunnel junctions appear particularly well suited for the implementation of electron heat rectifiers. Heat transport in such structures is deeply influenced by the strong temperature dependence of the superconducting density of states (DOS). Yet, the Josephson effect provides the thermal diode with an even more interesting capability. As recently demonstrated,\cite{Giazottoarxiv} heat currents flowing through Josephson tunnel junctions depend also on the macroscopic quantum phase difference of the Cooper pair condensates,\cite{MakiGriffin,Guttman97,Zhao03} just as charge currents do.  This latter property has outstanding consequences enabling to conceive thermal quantum devices that go well beyond the simple concept of heat rectification. All these features lead, under suitable conditions, to remarkable rectification coefficients as large as $\sim 800\%$.    


\begin{figure}[t]
\includegraphics[width=\columnwidth]{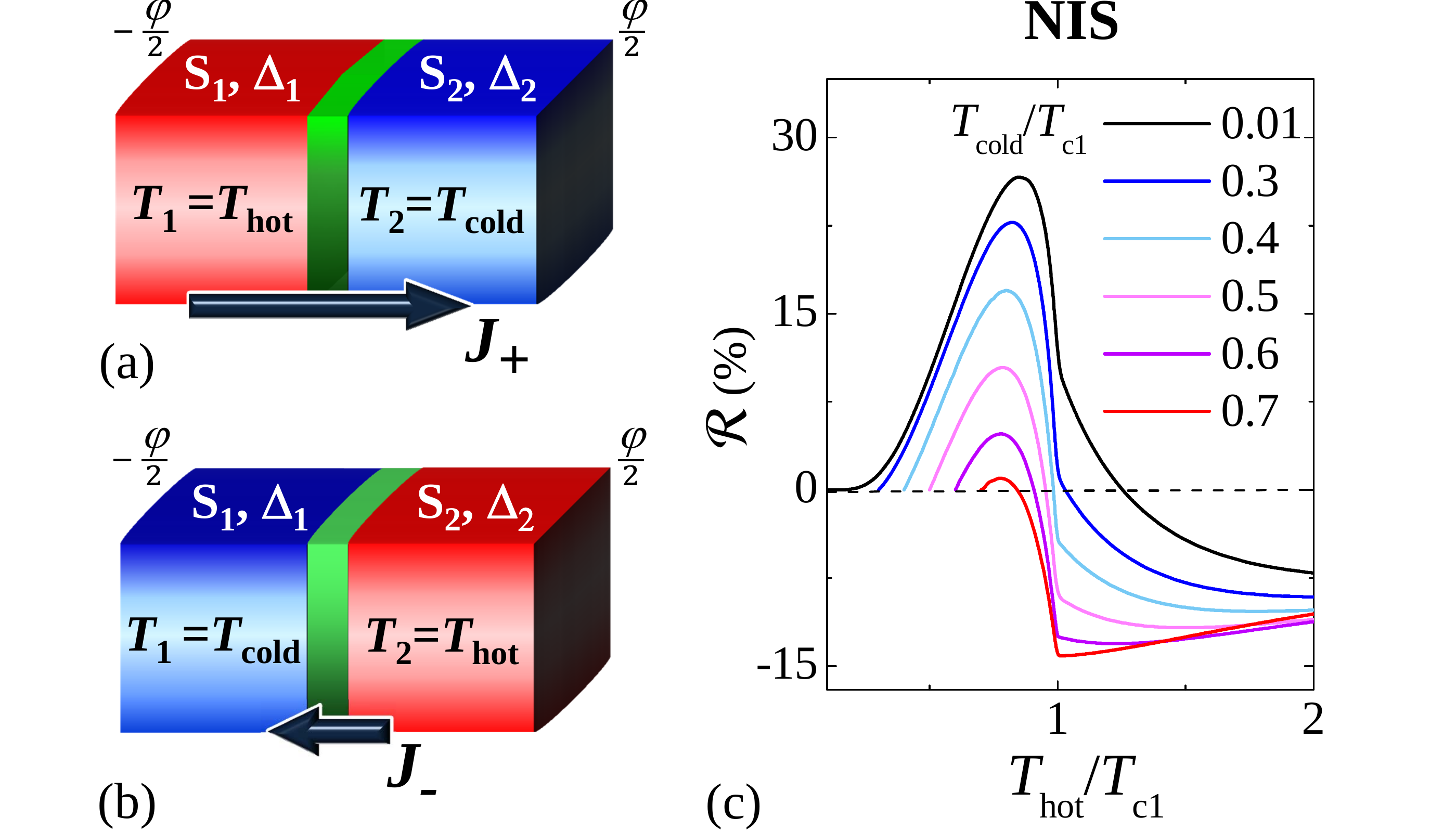}
\caption{(a) and (b) Josephson thermal diode scheme corresponding to the forward and reverse thermal bias configuration, respectively. The diode is phase biased ($\varphi$) and temperature biased with $T_1 \neq T_2$  as well, but the voltage across the junction vanishes. (c) $\mathcal{R}$ vs. $T_{\textrm{hot}}$ for different values of $T_{\textrm{cold}}$ corresponding to a normal metal/insulator/superconductor (NIS) thermal diode so that $\Delta_2 =0$.}
\label{Fig1}
\end{figure}

We shall start, first of all, by defining a heat rectification parameter $\mathcal{R}$. To this end, let us consider two different superconductors, S$_1$ and S$_2$, weakly coupled so to implement a Josephson tunnel junction as shown in Fig. \ref{Fig1}(a) and \ref{Fig1}(b). Each superconductor is characterized by its energy gap $\Delta_1$  and $\Delta_2$ leading to critical temperatures $T_{\textrm{c1}}$ and $T_{\textrm{c2}}$, respectively. The electronic temperature in both S$_1$ and S$_2$ is kept at fixed  $T_1$  and $T_2$, respectively, and the voltage drop across the junction is set to zero.  Additionally, $\varphi$ denotes the macroscopic phase difference across the junction with normal-state resistance $R_{\textrm{J}}$. In the forward thermal bias configuration, a thermal gradient is created by setting $T_1 = T_{\textrm{hot}}  > T_2= T_{\textrm{cold}}$ which leads to a total heat current $J_{+}$ flowing from S$_1$ to S$_2$ [see Fig. \ref{Fig1}(a)]. In the reverse thermal bias configuration, the thermal gradient is inverted so that $T_1= T_{\textrm{cold}}< T_2= T_{\textrm{hot}}$  leading to a heat current  $J_{-}$ flowing from S$_2$ to S$_1$ [see Fig. \ref{Fig1}(b)]. Under these hypothesis we define the rectification coefficient as
\begin{equation}
\mathcal{R} (\%)= \frac{J_{+}-J_{-}}{J_{-}}\times100. 
\end{equation}

We describe now the equations governing heat transport in the Josephson thermal rectifier. We focus on the electronic contribution to heat transport only, and neglect eventual heat currents carried by lattice phonons.  The forward and reverse total heat currents flowing through the Josephson junction read\cite{GiazottoAPL12}
\begin{eqnarray}
\begin{aligned}
 J_{+(-)}= J_{qp}[ T_{\textrm{hot(cold)}},& T_{\textrm{cold(hot)}}] \\
 -J_{int}[&T_{\textrm{hot(cold)}}, T_{\textrm{cold(hot)}}]\cos\varphi, \label{forward} 
\end{aligned}
\end{eqnarray}
where the term $J_{qp}$ accounts for the energy carried by quasiparticles,\cite{MakiGriffin} 
\begin{eqnarray}
\begin{aligned}
J_{qp}(T_k, T_l) = \frac{ 2 }{e^2 R_{\textrm{J}} } \int_0 ^\infty \varepsilon  \mathcal{N}_1(\varepsilon , T_{k})  &  \mathcal{N}_2  (\varepsilon , T_{l}) \\ \times & [f(T_{\textrm{hot}}) - f(T_{\textrm{cold}}) ]d \varepsilon,
\label{quasiparticles} 
\end{aligned}
\end{eqnarray}
and $k,l=$hot,cold. In Eq. (\ref{quasiparticles}), $ \mathcal{N}_{\alpha}(\varepsilon , T_k) = \Re  \left[ \frac{ \varepsilon/\Delta_{\alpha}(T_{k})+i \gamma}{  \sqrt{[\varepsilon/\Delta_{\alpha}(T_{k})+i \gamma]^2- 1}}  \right]   $ is the normalized smeared (by non zero $\gamma$) BCS quasiparticle DOS in  S$_{\alpha}$ with $\alpha=1,2$.\cite{Dynes} In the following we will assume $\gamma \sim   10^{-5} $, which describes realistic SIS' junctions.\cite{pekola2,pekola,giazotto04} Additionally, $ f(T_k)=  (1+e^{\varepsilon /  k_{\texttt{B}} T_{k}})^{-1}$ is the Fermi energy distribution, $\Delta_{\alpha}(T_{k})$ is the temperature-dependent energy gap of S$_{\alpha}$, $k_{\texttt{B}}$ is the Boltzmann constant and $e$ is the electron charge. On the other hand, the term $J_{int}$ in Eq. (\ref{forward}) is the amplitude of the phase-dependent component of the heat current,\cite{MakiGriffin,Guttman97,Zhao03} 
\begin{eqnarray}
\begin{aligned}
J_{int}(T_k, T_l) = \frac{ 2}{e^2 R_{\textrm{J}} }    \int _0 ^\infty\varepsilon   \mathcal{M}_1(\varepsilon , T_{k}) &\mathcal{M}_2   (\varepsilon , T_{l}) \\ \times & [f(T_{\textrm{hot}}) - f(T_{\textrm{cold}}) ]d \varepsilon,
\label{int} 
\end{aligned}
\end{eqnarray}
where  $\mathcal{M}_{\alpha}(\varepsilon , T_k) = \Im  \left[ \frac{1}{i \sqrt{[\varepsilon/\Delta_{\alpha}(T_{k})+i \gamma]^2- 1}}  \right]$.\cite{Bergeret} This component is peculiar to the Josephson effect and arises as a consequence of tunneling processes  through the junction involving both quasiparticles and Cooper pairs.\cite{MakiGriffin,Guttman97,Zhao03} It is worthwhile to emphasize that, depending on the value of  $\varphi$,  the second term in Eq. (\ref{forward})  may switch its sign.  The existence and sign of this component was experimentally demonstrated in Ref.\citenum{Giazottoarxiv} 

\begin{figure}[t]
\includegraphics[width=\columnwidth]{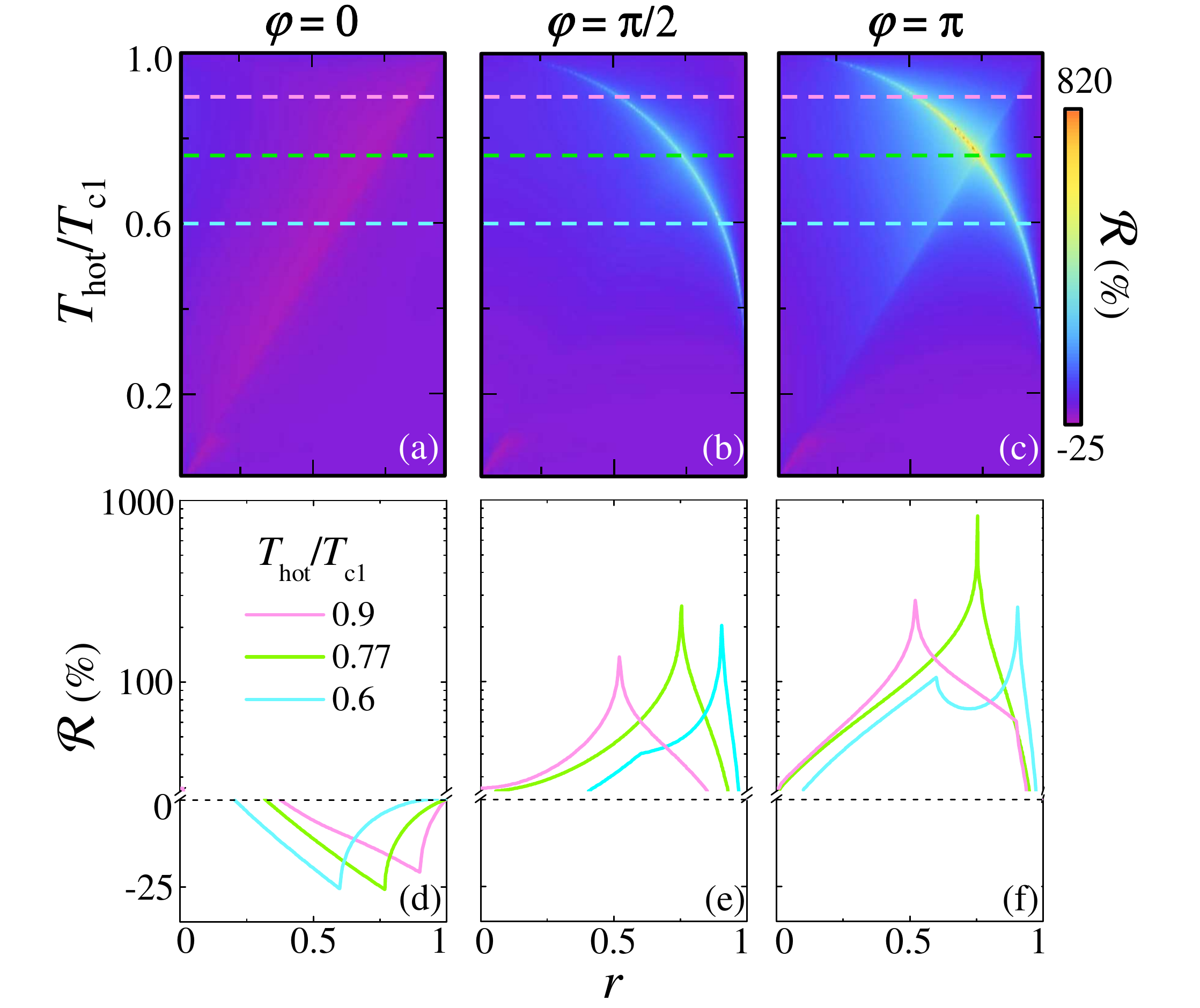}
\caption{ Panels (a), (b) and (c) show three density plots of $\mathcal{R}$  vs. $T_{\textrm{hot}}$ and $r$ calculated for $\varphi = 0$, $\varphi = \pi / 2$ and $\varphi = \pi$, respectively.  Panels (d), (e) and (f) show three selected profiles of $\mathcal{R}$ vs $r$ for the same values of $\varphi$ corresponding to the colored dotted lines in (a), (b) and (c). Notice that the scale is logarithmic above the break in the vertical axis. In addition, a dashed line indicates $\mathcal{R}=0$.   All curves have been calculated for $T_{\textrm{cold}} = 0.01 T_{\textrm{c1}}$. }
\label{Fig2}
\end{figure}

It is illustrative to start by analyzing the case in which one of the two electrodes is a normal metal, i.e., a NIS junction. For this purpose we can simply set $\Delta_2 = 0$ which leads to the complete suppression of 
$J_{int}$ in Eq. (\ref{forward}).  We calculate $\mathcal{R}$ as a function of $T_{\textrm{hot}}$ for different values of $T_{\textrm{cold}}$, i.e., for different temperature gradients established across the weak link. As shown in Fig.  \ref{Fig1}(c) a maximum positive rectification of $\sim 26 \% $ is obtained for $T_{\textrm{hot}} \simeq 0.85 T_{\textrm{c1}} $ at $T_{\textrm{cold}} = 0.01 T_{\textrm{c1}} $. As $T_{\textrm{hot}}$ increases, heat rectification starts to decrease eventually inverting its sign which implies that heat flux from the normal metal to S$_1$ becomes preferred. Furthermore, by increasing $T_{\textrm{cold}}$ leads to a reduction of $\mathcal{R}$ which reaches its maximum for larger values of $T_{\textrm{hot}}$. We note than, a bare NIS junction offers, in this way, an already quite rich response in terms of heat rectification.

In the following analysis we shall focus on the case in which both junction electrodes are superconductors, i.e., a SIS' junction. We define for clarity $r= \Delta_1/ \Delta_2 \leq 1$. By fixing the temperature of the second electrode to $T_{\textrm{cold}} = 0.01 T_{\textrm{c1}}$, we calculate  $\mathcal{R}$ as a function of $T_{\textrm{hot}}$ and as a function of $r$. The result is plotted in Fig. \ref{Fig2}(a), (b) and (c) for three representative cases, corresponding to $\varphi = 0$, $\varphi = \pi / 2$ and $\varphi = \pi$, respectively. Three selected profiles of $\mathcal{R}$ as a function of $r$ for different $T_{\textrm{hot}}$ values are shown as well in Fig. \ref{Fig2}(d), (e) and (f) for the same values of $\varphi$.   $\mathcal{R}$  depends strongly on $r $ reaching its maximum  at $r\simeq 0.75$ and $T_{\textrm{hot}}\simeq0.77 T_{\textrm{c1}}$ for $\varphi = \pi$, dropping then to zero at $r = 1$.  The inspection of these graphs also reveals how phase biasing across the junction does make a substantial  difference. In particular, the heat rectification coefficient does not only change by almost two orders of magnitude from $\varphi = 0$ to $\varphi = \pi$ but it also switches its sign.  

\begin{figure}[t]
\includegraphics[width=\columnwidth]{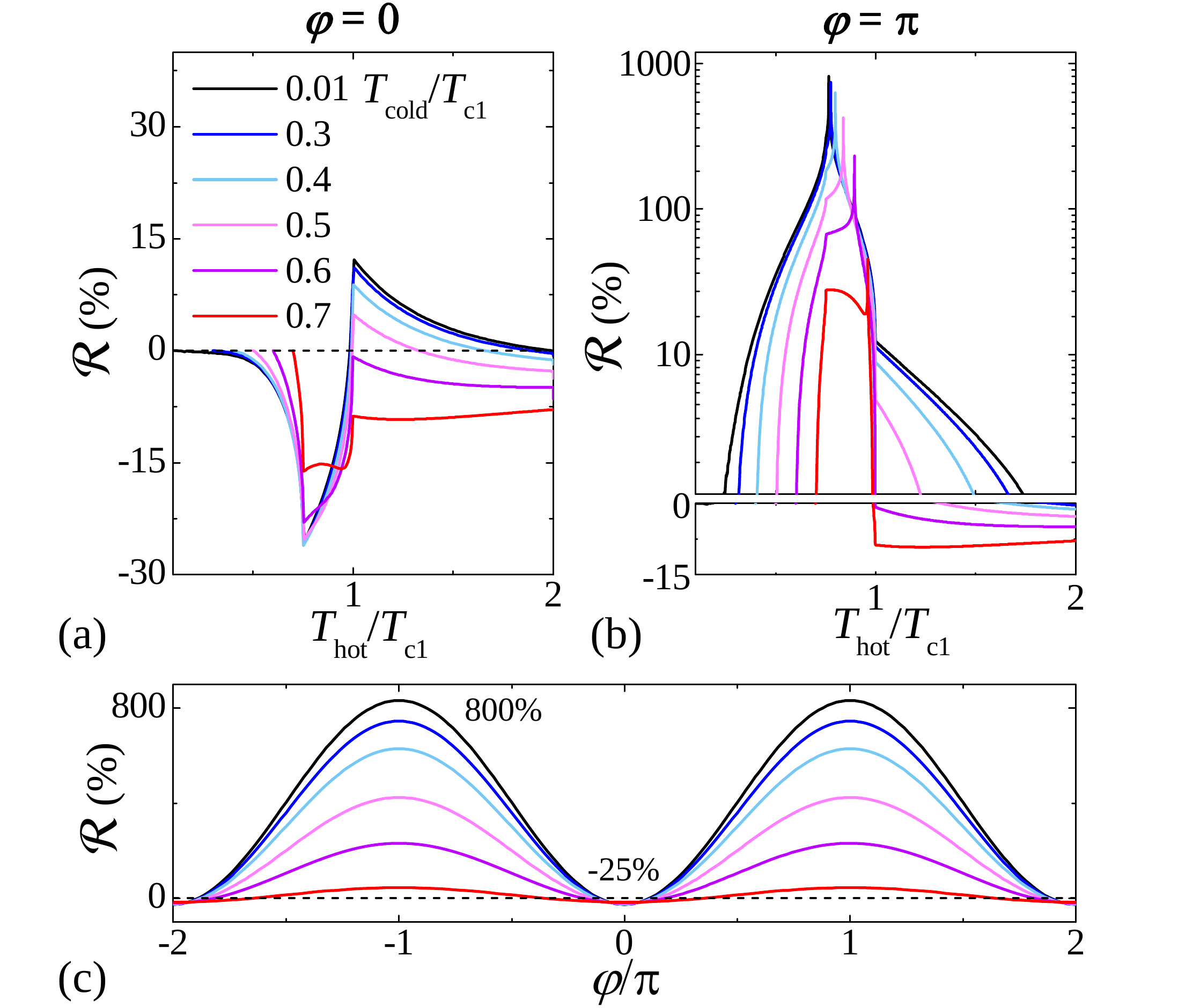}
\caption{(a) and (b) $\mathcal{R}$ vs. $T_{\textrm{hot}}$ for a few values of $T_{\textrm{cold}}$, at $\varphi = 0$ and $\varphi = \pi$, respectively. Notice that in (b) the vertical scale is logarithmic for $\mathcal{R}>0$. (c) Periodicity of $\mathcal{R}$ with $\varphi$ calculated for the same values of $T_{\textrm{cold}}$ and the corresponding $T_{\textrm{hot}}$ that maximize $\mathcal{R}$. The dashed line indicates $\mathcal{R}=0$. All these curves correspond to $r = 0.75$.}
\label{Fig3}
\end{figure}

The dependence of $\mathcal{R}$ on $T_{\textrm{hot}}$ for $\varphi = 0$ is shown in Fig. \ref{Fig3}(a) for different values of $T_{\textrm{cold}}$ and $r = 0.75$. In particular, $\mathcal{R}$ is negative as long as $T_{\textrm{hot}}<T_{\textrm{c1}}$, i.e., as long as one of the two electrodes remains in the superconducting state, reaching values of $\mathcal{R} \simeq - 25\%$ at $T_{\textrm{hot}} \simeq 0.76 T_{\textrm{c1}} $ and $T_{\textrm{cold}} = 0.01 T_{\textrm{c1}} $. Above this point, the sign of $\mathcal{R}$ and, therefore, the rectification direction depends on the thermal gradient being however slightly reduced. On the other hand, curves corresponding to $\varphi = \pi$ are shown in Fig. \ref{Fig3}(b) vs. $T_{\textrm{hot}}$ for the same values of $T_{\textrm{cold}}$. Remarkably, large values of $\mathcal{R} \sim 800 \%$  can be obtained  at $T_{\textrm{hot}} \simeq 0.77 T_{\textrm{c1}} $ and $T_{\textrm{cold}} = 0.01 T_{\textrm{c1}} $. If $T_{\textrm{cold}}$ is increased, $\mathcal{R}$ is in general reduced reaching its maximum for larger values of $T_{\textrm{hot}}$. Above $T_{\textrm{c1}}$, the behavior is identical to that calculated for $\varphi = 0$ since one electrode remains always in the normal state and the phase plays no role any more. In Fig. \ref{Fig3}(c), the periodicity of  $\mathcal{R}$ with $\varphi$ is shown for the same values of $T_{\textrm{cold}}$ and the corresponding $T_{\textrm{hot}}$ that maximize $\mathcal{R}$. We stress how $\varphi$-dependence provides the Josephson thermal diode with an unique tunability.

It is worthwhile to emphasize that the SIS' junction rectifies heat only if $ \Delta_1 \neq \Delta_2 $. As for the case of the NIS diode, heat rectification demands the combination of two different DOS  being (at least one of them) strongly temperature-dependent.\cite{Segal} Yet, a fairly large thermal gradient is required as well. Indeed, heat rectification is absent in the linear-response regime, that is, for small temperature differences $\delta T   = T_{\textrm{hot}} -T_{\textrm{cold}} << T = \frac{T_{\textrm{hot}}+T_{\textrm{cold}}}{2}$. In this case, the total electron heat current flowing through the  Josephson junction reduces to\cite{Bergeret}  
\begin{eqnarray}
\begin{aligned}
J_{\mathcal{L}}= &\frac{ \delta T}{2e^2 k_{\texttt{B}} T^2 R_{\textrm{J}} }  \int _0 ^\infty\varepsilon^2   \text{sech} ^2 \left( \frac{ \varepsilon}{2 k_{\texttt{B}} T }  \right)  \\ \times & \big[ \mathcal{N}_1(\varepsilon , T) \mathcal{N}_2   (\varepsilon , T)-   \mathcal{M}_1(\varepsilon , T) \mathcal{M}_2   (\varepsilon , T) \cos \varphi \big] d \varepsilon,  
\end{aligned}
\end{eqnarray}
which depends on the average temperature of the two electrodes only.

\begin{figure}[t]
\includegraphics[width=\columnwidth]{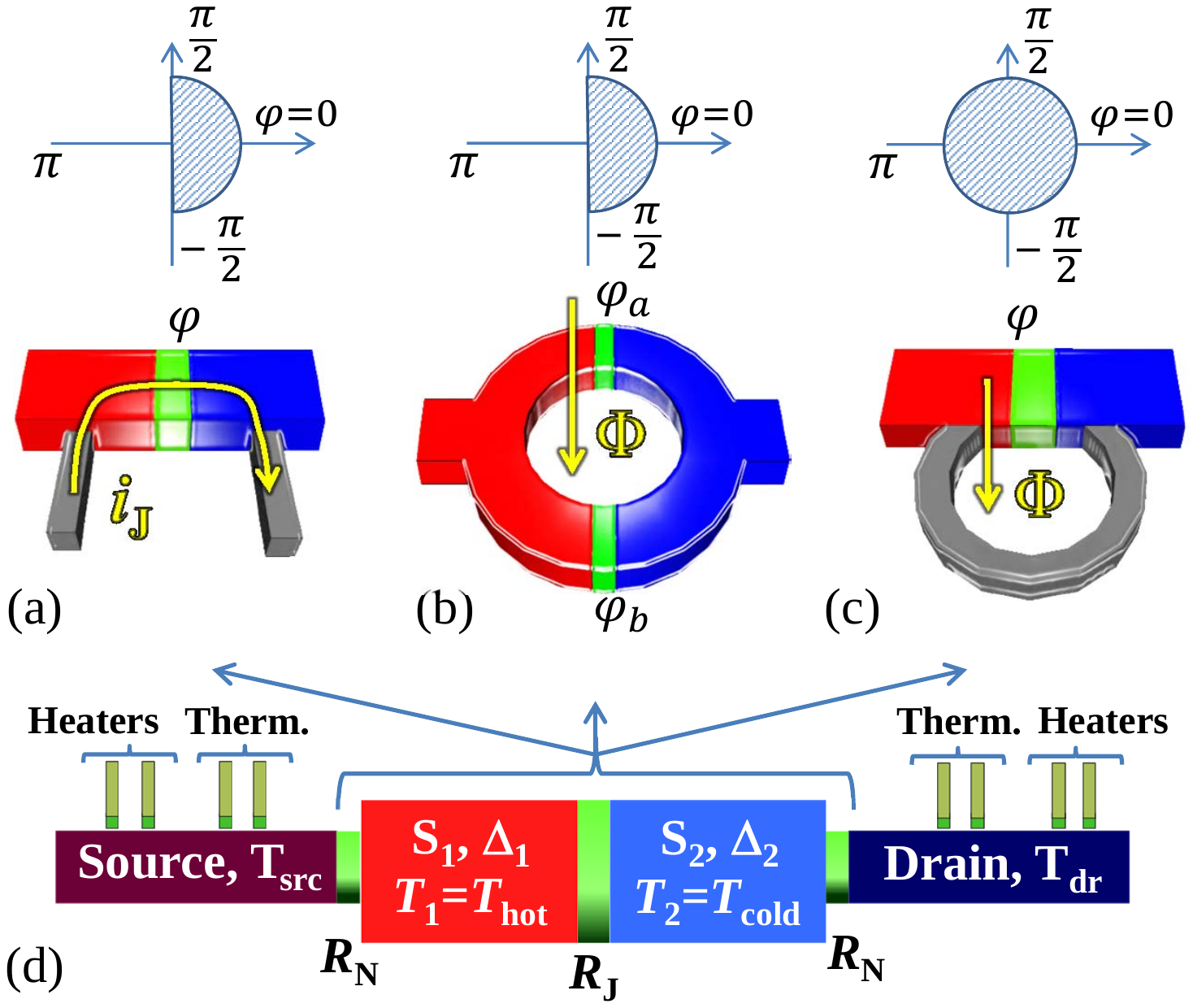}
\vspace*{-4.ex}
\caption{ Panels (a), (b) and (c) show three possible ways of  \itshape in-situ \upshape phase biasing of a Josephson junction within $ -\pi /2 \leq \varphi \leq \pi/2$ [(a) and (b)] or within $ -\pi \leq \varphi \leq \pi$ (c). (d) Possible experimental realization of the Josephson thermal rectifier. Source and drain normal metal electrodes are tunnel-coupled to the diode's core. Additional superconducting probes tunnel-coupled to source and drain allow for the implementation of SINIS thermometers and heaters.}
\label{Fig4}
\end{figure}

We discuss in the following some possible experimental realizations faced to make the most of the phase-dependence of the Josephson thermal rectifier. On the very first place we are interested in playing with the macroscopic phase difference across the Josephson junction during operation. Phase biasing of a Josephson junction can be achieved, in general, through supercurrent injection or by applying an external magnetic flux. In the former case, schematized in Fig. \ref{Fig4}(a), a Josephson current $i_{\textrm{J}}$ is forced to flow through the junction via two extra control superconducting wires connected to the diode's core through clean or tunnel contacts. The control wires can be made of a third superconductor S$_3$ with energy gap $\Delta_3 >> \Delta_1,\Delta_2$ so to suppress heat losses. The  \itshape phase-current \upshape relation under such circumstances is given by  $\sin(\varphi)= i_{\textrm{J}}/i_{\textrm{J}}^{\textrm{c}}$.\cite{Tinkham} Provided that $i_{\textrm{J}} \leq i_{\textrm{J}}^{\textrm{c}} $,  $i_{\textrm{J}}^{\textrm{c}} $ being the junction critical current, a phase gradient contained within the sections $ -\pi /2 \leq \varphi \leq \pi/2$ can be established. An  analogous phase gradient can be obtained using a direct current superconducting quantum interference device (DC SQUID) pierced by an external control magnetic flux $\Phi$ as shown in Fig. \ref{Fig4}(b). If both junctions are identical and neglecting the loop's geometrical inductance, the \itshape phase-flux \upshape relation must satisfy $\cos(\varphi_a) =\cos( \varphi_b )= \sqrt{[1-\cos(2 \pi \Phi/\Phi_0 )]/2}$ where $\varphi_a$ and $\varphi_b$ are the phase drops across each junction\cite{GiazottoAPL12,martinez} and $\Phi_0$ is the flux quantum.  The optimum phase configuration in terms of heat rectification, i.e., $ \varphi =\pi$, can be reached by using a rf SQUID as shown in Fig. \ref{Fig4}(c). Fur such a purpose, the thermal diode can be enclosed through clean contacts within a superconducting ring S$_3$ pierced by a control flux $\Phi$. Neglecting again the loop's inductance, the  phase-flux relation is given in this case by $\varphi = 2 \pi \Phi/\Phi_0 $\cite{Tinkham} enabling  the phase drop across the junction to vary within the whole phase space, i.e., $ -\pi  \leq \varphi \leq \pi$.

\begin{figure}[t]
\includegraphics[width=\columnwidth]{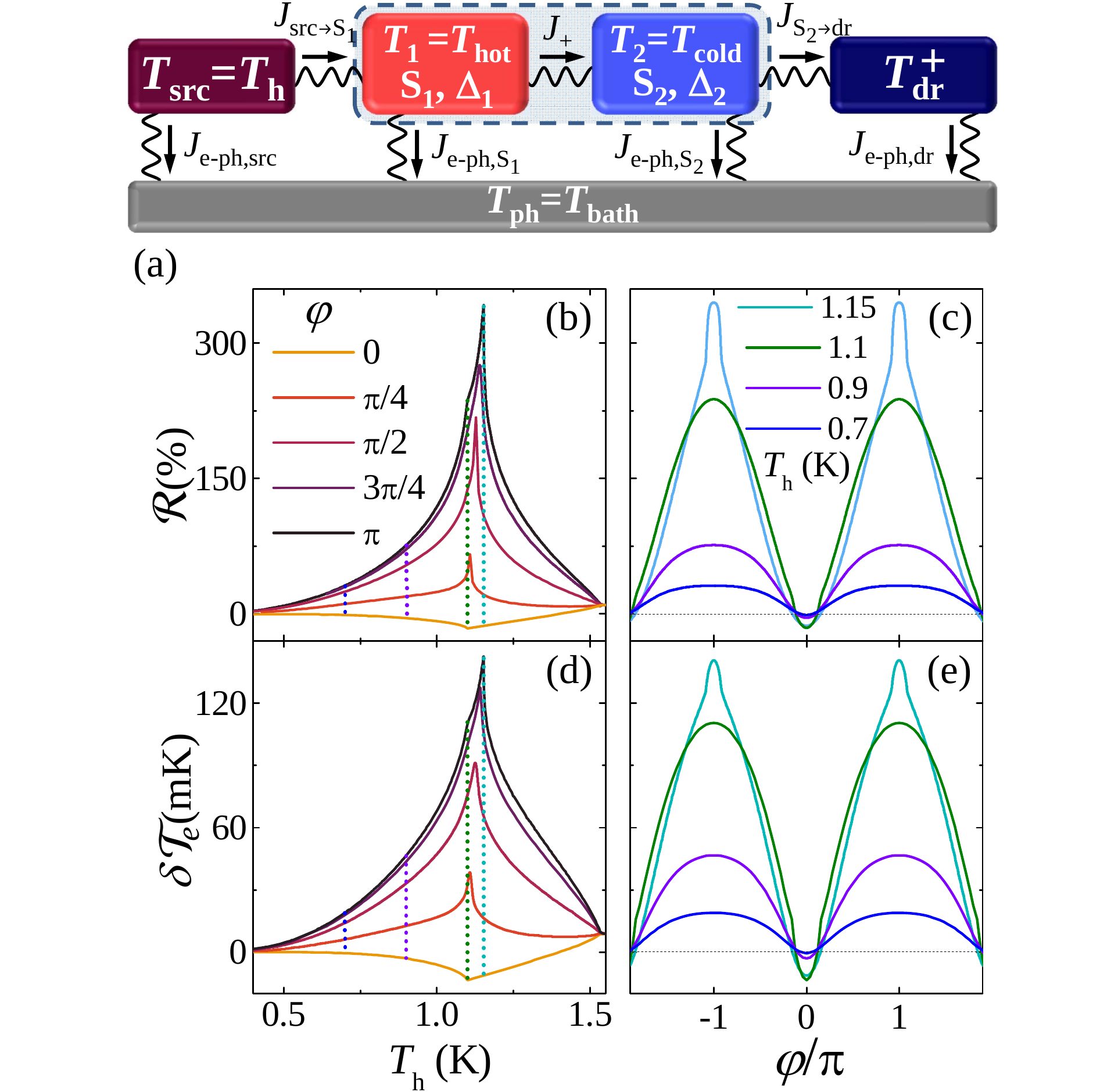}
\vspace*{-4.ex}
\caption{(a) Thermal model describing the main heat exchange mechanisms playing a role in the device proposed in Fig. \ref{Fig4}(d). Panels (b) and (c) show the computed values of $\mathcal{R} $ as a function of $T_{\textrm{h}}$ and $\varphi$, respectively. Dashed lines in (b) correspond to the curves in (c). Panels (d) and (e) show the same curves corresponding to $\delta \mathcal{T}_e $.}
\label{Fig5}
\end{figure}

Figure \ref{Fig4}(d) shows a device envisioned to probe experimentally the effects discussed above. Two identical normal metal electrodes, source and drain, are weakly connected one each via a resistance $R_{\textrm{N}}$ to both S$_1$ and S$_2$, respectively. Superconducting probes can be tunnel-coupled to these electrodes so to implement SINIS thermometers and heaters.\cite{GiazottoRev}  Yet, the forward thermal bias configuration can be realized by intentionally increasing the electronic temperature in source electrode up to $T_{\textrm{src}}^+ = T_{\textrm{h}} $ and probing the temperature in drain electrode $T_{\textrm{dr}}^+$. On the reverse configuration, we set $T_{\textrm{dr}}^- = T_{\textrm{h}} $ and  $T_{\textrm{src}}^-$ is measured in a similar way. The difference $\delta \mathcal{T}_e= T_{\textrm{dr}}^+ - T_{\textrm{src}}^-$ for a given $T_{\textrm{h}}$ can be used to assess  experimentally heat rectification. $\delta \mathcal{T}_e $ can be computed numerically by solving a system of thermal equations accounting for the heat exchange mechanisms present in our device [see Fig. \ref{Fig5}(a)]. On the forward configuration, electrons in S$_1$ exchange heat with electrons in the source at power $J_{\textrm{src}\rightarrow \textrm{S}_1}$ and, at power $J_{+}$,  with electrons in  S$_2$. On the other hand, electrons in S$_2$ exchange heat with drain's electrons at power $J_{ \textrm{S}_2 \rightarrow \textrm{dr}}$. Finally, electrons in the whole structure exchange heat at power $J_{\textrm{e-ph}}$ with lattice phonons that we assume to reside at bath temperature $T_{\textrm{bath}}$. Under such circumstances, the three unknown quantities, i.e., $T_{\textrm{hot}}$, $T_{\textrm{cold}}$ and $T_{\textrm{dr}}^+$ can be calculated for given initial conditions by solving the following system of thermal balance equations\cite{rev} 
\begin{eqnarray}
\begin{aligned}
 J_{\textrm{src}\rightarrow \textrm{S}_1} (T_{\textrm{h}},T_{\textrm{hot}})- J_{\textrm{e-ph,S}_1}(T_{\textrm{hot}}) -J_{+}(T_{\textrm{hot}},T_{\textrm{cold}})= 0, \\ 
 J_{+}(T_{\textrm{hot}},T_{\textrm{cold}}) - J_{\textrm{e-ph,S}_2}(T_{\textrm{cold}})- J_{ \textrm{S}_2 \rightarrow \textrm{dr}}(T_{\textrm{cold}},T_{\textrm{dr}}^+)= 0,    \\ 
 J_{ \textrm{S}_2 \rightarrow \textrm{dr}}(T_{\textrm{cold}},T_{\textrm{dr}}^+)-J_{\textrm{e-ph,dr}}(T_{\textrm{bath}},T_{\textrm{dr}}^+)= 0.  \end{aligned}   
\end{eqnarray}
In the above expressions, $J_{\textrm{e-ph,src(dr)}} = \Sigma_{\textrm{N}} \mathcal{V}_{\textrm{N}} (T_{\textrm{src(dr)}}^{5}-T_{\textrm{bath}}^{5})$\cite{GiazottoRev}, $\mathcal{V}_{\textrm{N}} $ and $\Sigma_{\textrm{N}} $ being the volume of the normal metal electrode and the electron-phonon coupling constant, respectively. Furthermore, we assume $T_{\textrm{bath}} = 10 $ mK $ << T_{k}  <<\Delta (T_k) / k_{\textrm{B}} $ so that  $J_{\textrm{e-ph,S}_{\alpha}} \simeq 0.95 \Sigma_{\textrm{S}}  \mathcal{V}_{\textrm{S}}  T_k^5 e^{ \frac{-\Delta_{\alpha} (T_k) }{ k_{\textrm{B}} T_k}}$,\cite{Timofeev} with $\mathcal{V}_{\textrm{S}} $ and $\Sigma_{\textrm{S}} $ being the volume of each superconducting electrode and the electron-phonon coupling constant, respectively.  As representative  parameters we set $\mathcal{V}_{\textrm{N}}=\mathcal{V}_{\textrm{S}} = 2 \times 10^{-20}$ m$^{3}$, $R_{\textrm{J}} = 10$ k$\Omega$ and $R_{\textrm{N}} = 100$ $\Omega$. Source and drain electrodes can be made, for instance, of Cu for which  $\Sigma_{\textrm{N}}\simeq 3 \times 10^9$ WK$^{-5}$m$^{-3}$\cite{GiazottoRev} whereas the diode's core can be made of Al and Mn-doped Al with $\Sigma_{\textrm{S}} \simeq 0.3 \times 10^9$ WK$^{-5}$m$^{-3}$,\cite{GiazottoRev} since the latter allows for fine tuning of the aluminum superconducting gap.\cite{Oneil,Oneil2} In this way, we set $T_{\textrm{c1}} = \frac{\Delta_1}{1.764 k_{\textrm{B}} }= 1.4$ K and $r=0.75$. With this set of parameters, $\mathcal{R}$ can be determined as a function of $T_{\textrm{h}}$ for given values of $\varphi$. The resulting curves are  plotted in Fig. \ref{Fig4}(b) together with the computed values of  $\delta \mathcal{T}_e $ vs. $T_{\textrm{h}}$ which are plotted in Fig. \ref{Fig4}(d). Remarkably, in the present setup,  a maximum $ \mathcal{R} \sim 340 \%$ can be reached. The latter corresponds to a temperature difference as large as $\delta \mathcal{T}_e  \sim 140$ mK which is easily measurable with standard SINIS or SNS thermometry techniques.\cite{GiazottoRev,Giazottoarxiv} Even more interesting, phase-coherence fingerprints are clearly observable as well. Notably, $\mathcal{R}$ and $\delta \mathcal{T}_e$ show the expected $2\pi$-periodicity as shown in Fig. \ref{Fig4}(c) and (e). In closing, we emphasize that, at such low bath temperatures, both superconducting electrodes are only marginally coupled to phonons.\cite{GiazottoRev} Indeed, neglecting the contribution of $J_{\textrm{e-ph,S}_i} $ leads to differences less than $\sim 5 \%$ of the values presented here. 

To summarize, we have proposed and analyzed the concept of a Josephson thermal rectifier. Under appropriate conditions, a remarkably large rectification coefficient of $\mathcal{R} \sim 800 \%$ can be obtained.  In addition, the Josephson thermal diode is phase-tunable. This latter property allows to maximize heat rectification \emph{ in-situ} or, even, to switch its sign. Such a device might find a straightforward  application, e.g., in the field of electronic refrigeration enabling magnetic-flux dependent heat management and thermal isolation at the nanoscale.  The operation principle which is at the basis of this heat rectifier will likely contribute, on the other hand, to improve the performance of other different coherent thermal components such as heat transistors or splitters.\cite{Giazottoarxiv,GiazottoAPL12,martinez} These thermal devices might potentially lead to the emergence of coherent caloritronic nanocircuits.


We acknowledge R. Aguado and C. Altimiras for comments, and the FP7 program No. 228464 "MICROKELVIN", the Italian Ministry of Defense through the PNRM project "TERASUPER", and the Marie Curie Initial Training Action (ITN) Q-NET 264034 for partial financial support.




\begin{thebibliography}{99}
\bibitem{GiazottoRev} F. Giazotto, T. T. Heikkil\"{a}, A. Luukanen, A. M. Savin, and J. P. Pekola, Rev. Mod. Phys. \textbf{78}, 217 (2006).
\bibitem{Dubi} Y. Dubi and M. Di Ventra, Rev. Mod. Phys. \textbf{83}, 131 (2011).
\bibitem{LiRev} N. Li, J. Ren, L. Wang, G. Zhang, P. H\"{a}nggi, and B. Li, Rev. Mod. Phys. \textbf{84}, 1045 (2012).
\bibitem{NielsenChuang} M. A. Nielsen and I. L. Chuang, Quantum Computation and Quantum Information (Cambridge University Press, 2002).
\bibitem{Giazottoarxiv} F. Giazotto and M. J. Mart\'inez-P\'erez, Nature \textbf{492}, 401 (2012).
\bibitem{GiazottoAPL12} F. Giazotto and M. J. Mart\'inez-P\'erez, Appl. Phys. Lett. \textbf{101}, 102601 (2012).
\bibitem{martinez} M. J. Mart\'inez-P\'erez and F. Giazotto, Appl. Phys. Lett. \textbf{102}, 092602 (2013).
\bibitem{RobertsRev} N.A. Roberts, D.G. Walker, Int. J. Therm. Sci., \textbf{50}, 648 (2011).
\bibitem{CasatiNandV} G. Casati, Nature Nanotech. \textbf{2}, 23 (2007).
\bibitem{Segal} L.-A. Wu and D. Segal, Phys. Rev. Lett. \textbf{102}, 095503 (2009). 
\bibitem{Segal2} D. Segal, Phys. Rev. Lett. \textbf{100}, 105901 (2008). 
\bibitem{Li} B. Li, L. Wang, G. Casati,  Appl. Phys. Lett. \textbf{88} (2006).
\bibitem{Terraneo} M. Terraneo, M. Peyrard, and G. Casati, Phys. Rev. Lett. \textbf{88}, 094302 (2002).
\bibitem{ren} J. Ren and J.-X. Zhu, Phys. Rev. B \textbf{87}, 165121 (2013).  
\bibitem{Ruokola1}  T. Ruokola and T. Ojanen,Phys. Rev. B \textbf{83}, 241404 (2011).
\bibitem{Kuo} D.M.T. Kuo, Y.C. Chang, Phys. Rev. B \textbf{81}, 205321 (2010).
\bibitem{Ruokola2} T. Ruokola, T. Ojanen, A.-P. Jauho,Phys. Rev. B  \textbf{79}, 144306 (2009).
\bibitem{Chen} X.-O. Chen, B. Dong, X.-L. Lei, Chin.Phys.Lett. \textbf{25}, 8 (2008).
\bibitem{Scheibner} R. Scheibner, M. K\"{o}nig, D. Reuter, A. D. Wieck, C. Gould, H. Buhmann, and L. W. Molenkamp, New J. Phys. \textbf{10}, 083016
(2008).
\bibitem{Chang} C. W. Chang, D. Okawa, A. Majumdar, and A. Zettl, Science \textbf{314}, 1121 (2006).
\bibitem{MakiGriffin} K. Maki and A. Griffin, Phys. Rev. Lett. \textbf{15}, 921 (1965).
\bibitem{Guttman97} G. D. Guttman, B. Nathanson, E. Ben-Jacob, and D. J. Bergman, Phys. Rev. B \textbf{55}, 3849 (1997). 
\bibitem{Zhao03} E. Zhao, T. L\"{o}fwander, and J. A. Sauls, Phys. Rev. Lett. \textbf{91}, 077003 (2003).
\bibitem{Dynes} R. C. Dynes, V. Narayanamurty and J. P. Garno, Phys. Rev. Lett. \textbf{41}, 1509 (1978).
\bibitem{pekola2} O.-P. Saira, A. Kemppinen, V. F. Maisi, and J. P. Pekola, Phys. Rev. B \textbf{85}, 012504 (2012).
\bibitem{pekola} J. P. Pekola, V. F. Maisi, S. Kafanov, N. Chekurov, A. Kemppinen, Yu. A. Pashkin, O.-P. Saira, M. M\"{o}tt\"{o}nen, and J. S. Tsai, Phys. Rev. Lett. \textbf{105}, 026803 (2010). 
\bibitem{giazotto04} J. P. Pekola, T. T. Heikkil\"{a}, A. M. Savin, J. T. Flyktman, F. Giazotto, and F. W. J. Hekking, Phys. Rev. Lett.  \textbf{92}, 056804 (2004).
\bibitem{Bergeret} F. Giazotto, and F. S. Bergeret, Appl. Phys. Lett. \textbf{102}, 132603 (2013).
\bibitem{Tinkham} M. Tinkham, Introduction to Superconductivity, (McGraw-Hill, 1996) and references therein.
\bibitem{rev} $T_{\textrm{src}}^-$ can be obtained in a similar way for the reverse configuration by simply exchanging the roles of $J_{\textrm{src}\rightarrow \textrm{S}_1} \Leftrightarrow J_{\textrm{dr}\rightarrow \textrm{S}_2} $, $J_{\textrm{e-ph,S}_1} \Leftrightarrow J_{\textrm{e-ph,S}_2}$, $J_{+} \Leftrightarrow J_{-}$, $J_{ \textrm{S}_2 \rightarrow \textrm{dr}} \Leftrightarrow J_{ \textrm{S}_1 \rightarrow \textrm{src}}$ and $J_{\textrm{e-ph,dr}} \Leftrightarrow J_{\textrm{e-ph,src}}$. $J_{\textrm{src} \leftrightarrow \textrm{S}_i}$ can be obtained from Eq. (\ref{int}) by setting $\Delta_i(T_{k}) = 0$ for the normal metal.
\bibitem{Timofeev} A. V. Timofeev, C. Pascual Garcia, N. B. Kopnin, A. M. Savin, M. Meschke, F. Giazotto, and J. P. Pekola, Phys. Rev. Lett. \textbf{102}, 017003 (2009).
\bibitem{Oneil} G. O'Neil, D. Schmidt, N. A. Miller,J. N. Ullom, A. Williams, G. B. Arnold and S. T. Ruggiero, Phys. Rev. Lett. \textbf{100}, 056804 (2008). 
\bibitem{Oneil2} G. O'Neil, D. Schmidt, N. A. Tomlin and J. N. Ullom, J. Appl. Phys. \textbf{107}, 093903 (2010).



\end{thebibliography}
\end{document}